 \documentclass[smallabstract,smallcaptions]{dccpaper}

\usepackage{epsfig}
\usepackage{citesort}
\usepackage{amsmath}
\usepackage{amssymb}
\usepackage{color}
\usepackage{url}
\def\@fnsymbol#1{\ensuremath{\ifcase#1\or *\or \dagger\or \ddagger\or
   \mathsection\or \mathparagraph\or \|\or **\or \dagger\dagger
   \or \ddagger\ddagger \else\@ctrerr\fi}}
\newcommand{\ssymbol}[1]{^{\@fnsymbol{#1}}}

\newlength{\figurewidth}
\newlength{\smallfigurewidth}

\begin{document}

\title
{\large
\textbf{Compressed image quality assessment using stacking}
}

\author{%
S. Farhad Hosseini-Benvidi$^{\ast}$, Hossein Motamednia$^{\dag}$,\\ Azadeh Mansouri$^{\ast}$, Mohammadreza Raei$^{\ast}$ and Ahmad Mahmoudi-Aznaveh$\ssymbol{3}$ \\[0.5em]
{\small\begin{minipage}{\linewidth}\begin{center}
\begin{tabular}{c}
$^{\ast}$ \footnotesize{Department of} \footnotesize{Electrical and Computer Engineering,}\\
\footnotesize{Faculty of Engineering,}  \\
\footnotesize{Kharazmi University, Tehran, Iran} \\
\url{{farhad.hosseini,a_mansouri, mraei}@khu.ac.ir} 
\end{tabular}\\[0.5em]
\begin{tabular}{ccc}
$^{\dag}$\footnotesize{High Performance Computing Laboratory,} & \hspace*{0.5in} & $\ssymbol{3}$\footnotesize{Cyber Research Institute,} \\
\footnotesize{School of Computer Science,} && \footnotesize{Shahid Beheshti University} \\
\footnotesize{Institute for Research in Fundamental Sciences,} && \footnotesize{Tehran, Iran}\\
\footnotesize{Tehran, Iran} && \url{a_mahmoudi@sbu.ac.ir}\\
\url{h.motamednia@ipm.ir} && 
\end{tabular}
\end{center}\end{minipage}}
}

\maketitle

\thispagestyle{empty}

\begin{abstract}
It is well-known that there is no universal metric for image quality evaluation. In this case, distortion-specific metrics can be more reliable. The artifact imposed by image compression can be considered as a combination of various distortions. Depending on the image context, this combination can be different. As a result, Generalization can be regarded as the major challenge in compressed image quality assessment. In this approach, stacking is employed to provide a reliable method. Both semantic and low-level information are employed in the presented IQA to predict the human visual system. Moreover, the results of the Full-Reference (FR) and No-Reference (NR) models are aggregated to improve the proposed Full-Reference method for compressed image quality evaluation. The accuracy of the quality benchmark of the clic2024 perceptual image challenge was achieved 79.6\%, which illustrates the effectiveness of the proposed fusion-based approach.
\end{abstract}

\Section{Introduction}
With the rapid growth of digital communication, it has become more important to design a good image/video quality metric. A visual signal can be affected by a wide range of modifications/distortions.
These alterations may occur during signal acquisition, transmission, and compression. Image and video compression technologies are playing critical roles in a wide range of applications. The rate-distortion optimization is the heart of the lossy compression model for classic and learning-based approaches. As a result, developing a robust and accurate image and video quality metric is crucial, especially for image and video coding in the case of improving both the efficiency of image and video compression algorithms and enhancing the user experience in digital communication.

In many datasets such as TID2008\cite{tid2008}, TID2013\cite{tid2013}, LIVE\cite{live}, and CSIQ\cite{csiq}, the Mean Opinion Score(MOS) is considered the label of the distorted image. It is simpler to compare two input images and identify the one more similar to a reference instead of assigning quality scores. In \cite{prashnani2018pieapp},  a new, large-scale dataset is labeled with the probability that humans will prefer one image over another. Moreover, a deep-learning model is trained using a pairwise-learning framework to predict the preference. The trained network can then be used separately with only one distorted image and a reference to predict its perceptual error without ever being trained on explicit human Mean Opinion Score.
In fact, pairwise-learning framework trains
an error-estimation by the probability labels in dataset.
In the image perceptual model track of DCC 2024, the evaluation results for the image compression task should be correlated with the human visual system by proposing a metric. Given a pair of images (the original and the distorted image), the proposed metric should generate a score in which If $d(O, A) < d(O, B)$, the metric prefers method A over method B.

The distortion diversity and content variation are two major problems that appear in the task of authentic IQA. In \cite{su2020blindly} , to mimic the human visual system, the model separates quality prediction from content understanding. The hypernetwork architecture is proposed to capture complex distortions by introducing a multi-scale local distortion-aware module. 
NIQE, as a No-reference model, assesses the perceptual quality without knowledge of distortions or human opinions. A simple distance metric between the model statistics is employed to evaluate the distorted image's quality \cite{mittal2012making}.
These two methods are illustrations of semantic-based and low-level approaches.
Both semantic and low-level information are utilized in the proposed method to evaluate the results in two perspectives. In addition, the results of the Full-Reference (FR) and No-Reference (NR) models are combined to improve the compressed image quality evaluation.

\Section{Proposed Method}
The Human Visual System tends to extract structural information from viewing scenes. Structural information verification consists of two stages: semantic or content-based information and low-level features. Some methods, like SSIM and PSNR, were designed based on pixel-level features, and other approaches, such as LPIPS\cite{zhang2018unreasonable}, are more focused on extracting high-level semantic features. 
Based on this hypothesis, extracted features from the model, which is trained for object detection, can provide acceptable results for quality evaluations.

\begin{figure}[htbp]
  \centering
  \includegraphics[width=0.8\textwidth]{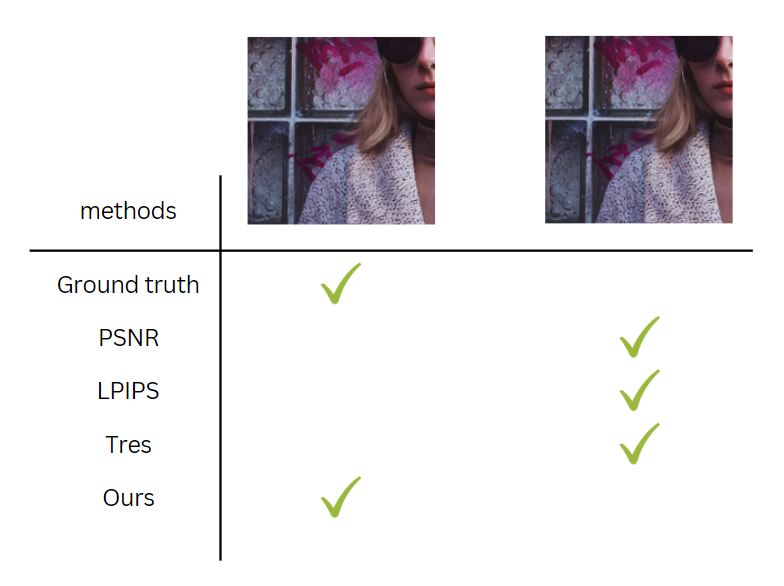}
  \caption{An example of a situation in the validation dataset where the conventional methods, namely PSNR, LPIPS, and Tres, selected the second image as the one more similar to the reference image. However, our method aligns more closely with human judgment and correctly identifies the first image as the better match.}
  \label{fig1:example}
\end{figure}

In figure \ref{fig1:example}, the prediction results for traditional approaches such as SSIM and one of the recent methods, LPIPS, illustrated their disagreement with human judgments. The presented fusion-based approach shows a better correlation with HVS.
At first attempts, we evaluated 15 image quality metrics, including No-reference and Full-reference methods employing the challenge dataset. The evaluation results are illustrated in Table 1.

\begin{figure}[htbp]
  \centering
  \includegraphics[width=1\textwidth]{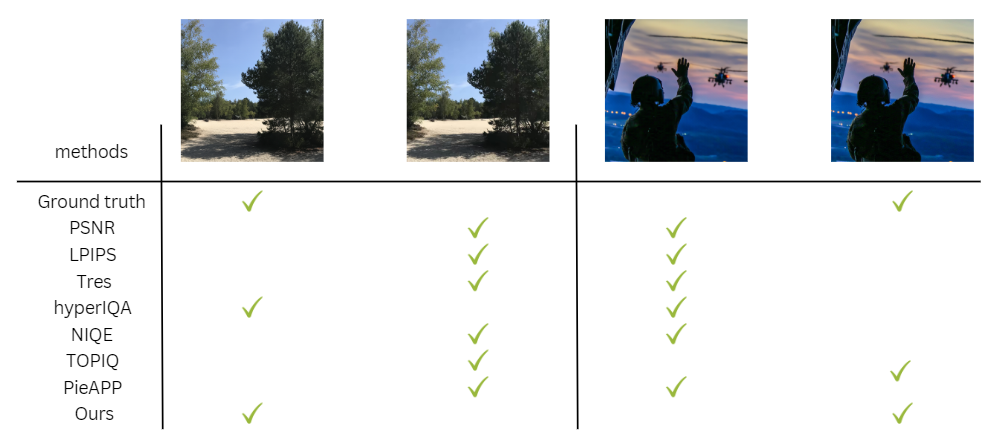}
  \caption{On the left side of the image, among the four methods chosen, three methods (NIQE, TOPIQ, and PieAPP) considered the second image superior in quality. However, our proposed method correctly identified the higher quality image with the assistance of a single vote from the hyperIQA method. The right side of the image presents a similar situation, but this time the TOPIQ method's vote was considered. This demonstrates that our method is not biased towards a specific method.}
  \label{fig2:example}
\end{figure}

In \cite{maniqa} Multi-dimension Attention
Network for no-reference Image Quality Assessment
(MANIQA) is presented. However, the results of the challenge dataset need to be better correlated with HVS.
In \cite{tres}, a hybrid combination of CNNs and Transformers features is extracted to show local and global features of the input image. Relative ranking and an additional self-consistency loss are considered to improve the robustness of
 the Tres. Although the experiments perform well in NR-IQA, Tres cannot provide acceptable results for the challenge dataset.
 In \cite{su2020blindly}, to mimic the human visual system, the model separates quality prediction from content understanding. The hypernetwork architecture is proposed to capture complex distortions by introducing a multi-scale local distortion-aware module. 
NIQE, as a No-reference model, assesses the perceptual quality without knowledge of distortions or human opinions. A simple distance metric between the model statistics is employed to evaluate the distorted image's quality \cite{mittal2012making}.
As illustrated in Table 1, both \cite{hyperiqa} and \cite{mittal2012making} cannot provide acceptable results in the challenge dataset.
A group of methods that consider pairwise learning, including PieAPP and LPIPS, achieve better performance. However, as we analyzed many samples in the dataset, the method failed to predict the results.
As illustrated in figure \ref{fig2:example}, two dataset samples are evaluated using various methods, including \cite{hyperiqa}, \cite{mittal2012making}, \cite{pieapp} and \cite{topiq}. The effectiveness of the proposed stacking-based approach is clearly shown.

In fact, each of the basic models can provide an evaluation considering different criteria.
One of the high-performance quality models is the fusion-based method, which explores models that present lower performances individually. One of the most prominent examples is VMAF, which was developed by Netflix and gained significant attention in video processing and content delivery \cite{li2018vmaf}.

\begin{figure}[htbp]
  \centering
  \includegraphics[width=0.6\textwidth]{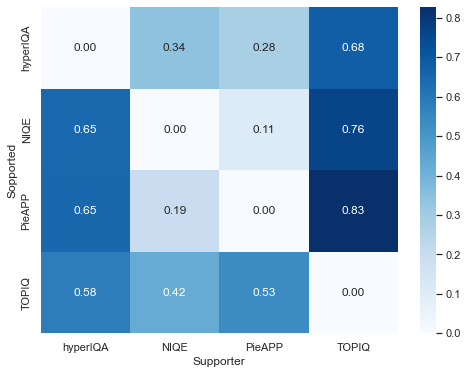}
  \caption{The supporter methods are applied just on supported False results}
  \label{fig3:example}
\end{figure}

In the proposed method, overall performance can be improved by combining the predictions of numerous base models \cite{breiman1996stacked}\cite{sharkey1996combining}\cite{wolpert1992stacked}. 
To explore the mutual effects of these methods, each metric's accuracy on others' false samples is illustrated in \ref{fig3:example}. This experiment shows that different methods cover each other to some extent.

\begin{figure}[htbp]
  \centering
  \includegraphics[width=0.8\textwidth]{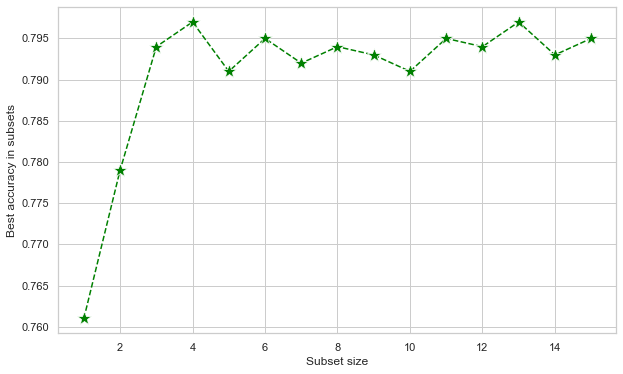}
  \caption{We conducted a thorough investigation on 15 quality assessment methods. By analyzing all the possible combinations of these 15 methods and representing the highest accuracy achieved by subsets of a fixed size as a scatter plot, we observed that the accuracy reaches a saturation when using subsets consisting of 4 methods. Additionally, we found that increasing the number of features does not lead to a significant improvement in accuracy.}
  \label{fig4:example}
\end{figure}

A feature vector is constructed using two models of size four(two values for each metric related to images A and B).
In another attempt, a feature vector of size six is generated to enhance the results by employing three metrics. Finally, the feature vector of size eight is created using four metrics. The top accuracy of each feature is illustrated in \ref{fig4:example}. 
SVM with RBF kernel is employed for all the built features to predict the quality label. The train and test cycle is performed five times, and the median of these five cycles is reported. We used
 80 percent of images for training and the remaining for testing. The outcomes of various combinations are illustrated in the figure \ref{fig4:example}.

 \begin{table}[htbp]
\centering
\caption{Methods and Accuracies}
\label{tab:methods-accuracies}
\begin{tabular}{|l|c|}
\hline
\textbf{Method} & \textbf{Accuracy (\%)} \\
\hline\hline
PSNR & 57.2 \\
CW-SSIM \cite{sampat2009complex}& 63.3 \\
STLPIPS-VGG \cite{ghildyal2022shift} & 76.1 \\
PieAPP \cite{prashnani2018pieapp}  & 75.4 \\
LPIPS-VGG \cite{zhang2018unreasonable} & 74.8 \\
LPIPS-alexNet \cite{zhang2018unreasonable} & 72.6 \\
TOPIQ(PIPAL) \cite{topiq} & 72.4 \\
NIQE \cite{mittal2012making}& 69 \\
MANIQ(PIPAL) \cite{maniqa} & 66.4 \\
hyperIQA \cite{hyperiqa}& 64.1 \\
IQA-CNN \cite{kang2014convolutional} & 61.8 \\
Tres \cite{tres}  & 61.3 \\
clipIQA+ vitL14 \cite{clipiqa}  & 61.3 \\
Tres(KONIQ) \cite{tres} & 60.7 \\
musiq(KONIQ) \cite{musiq} & 57.6 \\
\hline
\end{tabular}
\end{table}

 The best performance can be achieved using four basic methods: two No-reference and two Full-reference approaches. In another point of view, both semantic and low-level features are combined to make the final feature vectors. The experimental results illustrate the acceptable correlation in the challenge of image quality in CLIC2024.

\Section{Conclusion}
The IQA methods employ either semantic or low-level features. Although semantic features seem more robust for evaluating the effect of image compression, low-level features can be used to cover the weaknesses of high-level features. In addition, capturing the effect of various content using NR methods along with the FR ones can be appropriate. Comprehensive experiments using FR and NR and methods with various features  are conducted to show their pros and cons. Based on our analysis, we conclude that a combination of various methods can predict the HVS more accordingly. Four metrics are used (PiApp, NIQE, TOPIQ, and HyperIQA). SVM is used to achieve the final decision, which predicts 79.6\%  in accordance with HVS. 

\twocolumn

\onecolumn
  
\Section{References}
\bibliographystyle{IEEEbib}
\bibliography{refs}

\end{document}